\documentclass[sigconf]{acmart}
\usepackage{multirow}
\usepackage{array}

\usepackage{xcolor}

\AtBeginDocument{%
  }

\setcopyright{acmlicensed}
\copyrightyear{2025}
\acmConference[AIxCC]{Proceedings of the XX Conference}{2025}{}
\acmISBN{978-1-4503-XXXX-X/25/06}
\acmDOI{10.1145/nnnnnnn.nnnnnnn}

\begin{document}

\title{All You Need Is A Fuzzing Brain: An LLM-Powered System for Automated Vulnerability Detection and Patching}





\author{Ze Sheng}
\affiliation{%
  \institution{Texas A\&M University}
  \city{College Station}
  \country{US}
}
\email{zesheng@tamu.edu}

\author{Qingxiao Xu}
\affiliation{%
  \institution{Texas A\&M University}
  \city{College Station}
  \country{US}
}
\email{qingxiao@tamu.edu}

\author{Jianwei Huang}
\affiliation{%
  \institution{Texas A\&M University}
  \city{College Station}
  \country{US}
}
\email{jwhuang@tamu.edu}

\author{Matthew Woodcock}
\affiliation{%
  \institution{Texas A\&M University}
  \city{College Station}
  \country{US}
}
\email{matthewwoodc0@tamu.edu}

\author{Heqing Huang}
\affiliation{%
  \institution{City University of Hong Kong}
  \city{Hong Kong}
  \country{China}
}
\email{heqhuang@cityu.edu.hk}

\author{Alastair F. Donaldson}
\affiliation{%
  \institution{Imperial College London}
  \city{London}
  \country{UK}
}
\email{alastair.donaldson@imperial.ac.uk}

\author{Guofei Gu}
\affiliation{%
  \institution{Texas A\&M University}
  \city{College Station}
  \country{US}
}
\email{guofei@cse.tamu.edu}

\author{Jeff Huang}
\authornote{Team lead}
\affiliation{%
  \institution{Texas A\&M University}
  \city{College Station}
  \country{US}
}
\email{jeff@cse.tamu.edu}






\renewcommand{\shortauthors}{Ze et al.}

\begin{abstract}
Our team, \emph{All You Need Is A Fuzzing Brain}, was one of seven finalists in DARPA's Artificial Intelligence Cyber Challenge (AIxCC), placing fourth in the final round. 
During the competition, we developed a Cyber Reasoning System (CRS) that autonomously discovered 28 security vulnerabilities---including six previously unknown zero-days---in real-world open-source C and Java projects, and successfully patched 14 of them. 
The complete CRS is open source at \href{https://github.com/o2lab/afc-crs-all-you-need-is-a-fuzzing-brain}{\textcolor{blue!70!black}{github.com/o2lab/afc-crs-all-you-need-is-a-fuzzing-brain}}.

This paper provides a detailed technical description of our CRS, with an emphasis on its LLM-powered components and strategies. 
Building on AIxCC, we further introduce a public leaderboard for benchmarking state-of-the-art LLMs on vulnerability detection and patching tasks, derived from the AIxCC dataset. 
The leaderboard is available at \href{https://o2lab.github.io/FuzzingBrain-Leaderboard/}{\textcolor{blue!70!black}{o2lab.github.io/FuzzingBrain-Leaderboard}}.



\end{abstract}






\maketitle

\section{Background}
We first introduce some background information of AIxCC~\cite{aixcc}, which is necessary to understand the design choices of our CRS (which is called "{\em FuzzingBrain}" in the rest of this paper). Readers already familiar with AIxCC may go directly to Section~\ref{sec:overview}.

\begin{table*}[htbp]
\centering
\caption{AIxCC Tasks and Challenge Modes}
\label{tab:fuzzing_brain_overview}
\begin{tabular}{|l|l|}
\hline
\textbf{Component} & \textbf{Description} \\
\hline
\multicolumn{2}{|c|}{\textbf{Two Tasks}} \\
\hline
POV Generation & Generate binary exploit (.bin) to trigger vulnerability \\
Patch Generation & Generate diff patch to fix vulnerability \& pass functionality tests \\
\hline
\multicolumn{2}{|c|}{\textbf{Three Challenge Modes}} \\
\hline
Delta-Scan Mode & Input: Target commit | Detect commit-specific vulnerabilities \& Generate remediation \\
Full-Scan Mode & Input: Complete codebase | Full codebase vulnerability discovery \& remediation \\
SARIF Assessment Mode & Input: External reports (SARIF) | Validate vulnerability reports \\
\hline
\end{tabular}
\end{table*}

\subsection{POV and Patch Generation}

In AIxCC, participants are tasked with building autonomous \emph{vulnerability detection and patching} systems that operate effectively on real-world open-source projects. These systems must fulfill two critical requirements:  
(1) automatically generate \textit{Proofs-of-Vulnerability (POVs)}, and  
(2) produce \textit{patches} in the form of diff files that remediate the discovered issues.

\textbf{Proof-of-Vulnerability (POV).} Given a target software that contains one or more vulnerabilities (either seeded by AIxCC or zero-days), for each vulnerability, generate an input that triggers a sanitizer error when processed by a fuzzer harness.
AIxCC targets vulnerabilities in C and Java projects, and it uses \texttt{OSS-Fuzz}~\cite{ossfuzz}-compatible fuzzers, such as libFuzzer and AFL for C projects, and Jazzer for Java projects, to prove a vulnerability. 
For C projects, the software can be compiled using various sanitizers: {\tt AddressSanitizier}, {\tt MemorySanitizer}, and {\tt Undefined-BehaviourSanitizer}. 

The POV generation process can be viewed formally as follows:
%
\begin{align*}
& \texttt{ GeneratePOV(H, S)} \rightarrow (h, san, I) \\
\textit{s.t. } & \texttt{Trigger}(h, san, I, S) = \texttt{True}
\end{align*}

where $H$ represents a collection of fuzzer harnesses, $S$ is the target software, and $h \in H$ is a fuzzer harness, $san$ a sanitizer type, and $I$ a sanitizer-error triggering input.
The generated POV serves as concrete evidence of the vulnerability's exploitability and also provides a test case for patch validation.

\textbf{Patch Generation.} Given a target software $S$ which contains one or more vulnerabilities, for each vulnerability, generate a patch in the form of a diff file that satisfies the following requirements:
\begin{enumerate}

\item The patched software $S'$ compiles successfully (i.e., without syntax or build errors);

\item The patch eliminates all known proofs of the vulnerability:
$$\forall (h, san, I) \in \texttt{POVs}\; . \; \neg \texttt{Trigger}(h, san, I, S')$$
\noindent where \texttt{POVs} is a set of known POVs for the given vulnerability, $h \in H$ is a fuzzer harness, and $san$ is a sanitizer type;

\item The patch preserves functional correctness according to a given set \texttt{TestSuite} of regression tests for the software:
$$\forall T \in \texttt{TestSuite} \; . \; \texttt{Pass}(T, S')$$
\end{enumerate}

These requirements provide a degree of confidence that patches not only remove vulnerabilities but also preserve intended functionality, minimizing the introduction of regressions.

Importantly, when the patches submitted by a team are evaluated, the set \texttt{POVs} for a vulnerability is taken to be the union of valid proofs submitted for the vulnerability across \emph{all} teams.
Therefore, to be valid, it is not enough for a team's patch to simply eliminate the POVs discovered by the team; the patch must be general enough to work for POVs discovered by other teams as well.

Using the diff format for patches enables seamless integration with existing version control systems and open-source development workflows, facilitating review and maintainability.

\subsection{Three Challenge Modes}

To address diverse vulnerability management scenarios in real-world software development, AIxCC defines three distinct challenge modes that participating systems shall support: \textit{Delta-Scan}, \textit{Full-Scan}, and \textit{Static Analysis Report-Based (SARIF~\cite{sarif})}. Each mode corresponds to a different class of inputs and evaluation tasks.


\subsubsection{Delta-Scan Mode} 
This mode targets commit-based vulnerability analysis, focusing on changes introduced by a specific commit.  

\textbf{Input:} A target commit $C$, repository base state $S_{\text{base}}$ (state before applying commit $C$), source code repository $R$, and corresponding OSS-Fuzz fuzzer harnesses $H$. 



\subsubsection{Full-Scan Mode} 
This mode targets capabilities to discover vulnerabilities across the \textbf{entire codebase}, rather than restricting analysis to a single commit.  

\textbf{Input:}  A software state $S$ (specific version), source code repository $R$, and corresponding OSS-Fuzz fuzzer harnesses $H$. 



\subsubsection{SARIF Assessment Mode} 
This mode validates externally provided vulnerability reports, typically from static analysis tools or issue trackers. More details can be found in Section~\ref{sec:sarif}.

\textbf{Objective:} Confirm the accuracy of reported vulnerabilities and filter out false positives.  

\textbf{Input:} Structured vulnerability reports (e.g., SARIF) including affected functions, vulnerability classifications, location information, contextual metadata, etc.




\subsection{Competition Scoring}
The AIxCC scoring rule incentivizes submission accuracy and speed:
\[
Score = AM \times (VDS + PRS + SAS + BDL)
\]

\textbf{Accuracy Multiplier (AM):} 
\[
AM = 1 - \tfrac{1-r}{4}, \quad 
r = \tfrac{acc}{acc+inacc}
\]

\textbf{Time Multiplier:} 
\[
\tau = 0.5 + \tfrac{time_{rem}}{2 \times time_{window}}
\]

\textbf{POV (VDS):} $2 \times \tau$ if crash, else $0$  

\textbf{Patch (PRS):} $6 \times \tau$ if valid, else $0$  

\textbf{SARIF Assessment (SAS):} $1 \times \tau$ if correct, else $0$  

\textbf{Bundle Score (BDL):} $1 \times \tau$ if valid, else $0$ 

AIxCC had a total number of 60 challenges in the final round. Each challenge score is the sum of valid POV, patch, SARIF assessment, and bundle points, scaled by an accuracy multiplier.  
Each valid POV earns 2 points, patch 6 points, and SARIF assessment 1 point, all decaying over time to a 50\% minimum.  
In addition, bundle points (BDL) are awarded for grouped submissions. Details can be found in the AIxCC final scoring guide~\cite{aixcc-scoring-guide}.

\section{Overview of FuzzingBrain Architecture}\label{sec:overview}
\begin{figure*}[htbp]
    \centering
    \includegraphics[width=0.9\textwidth]{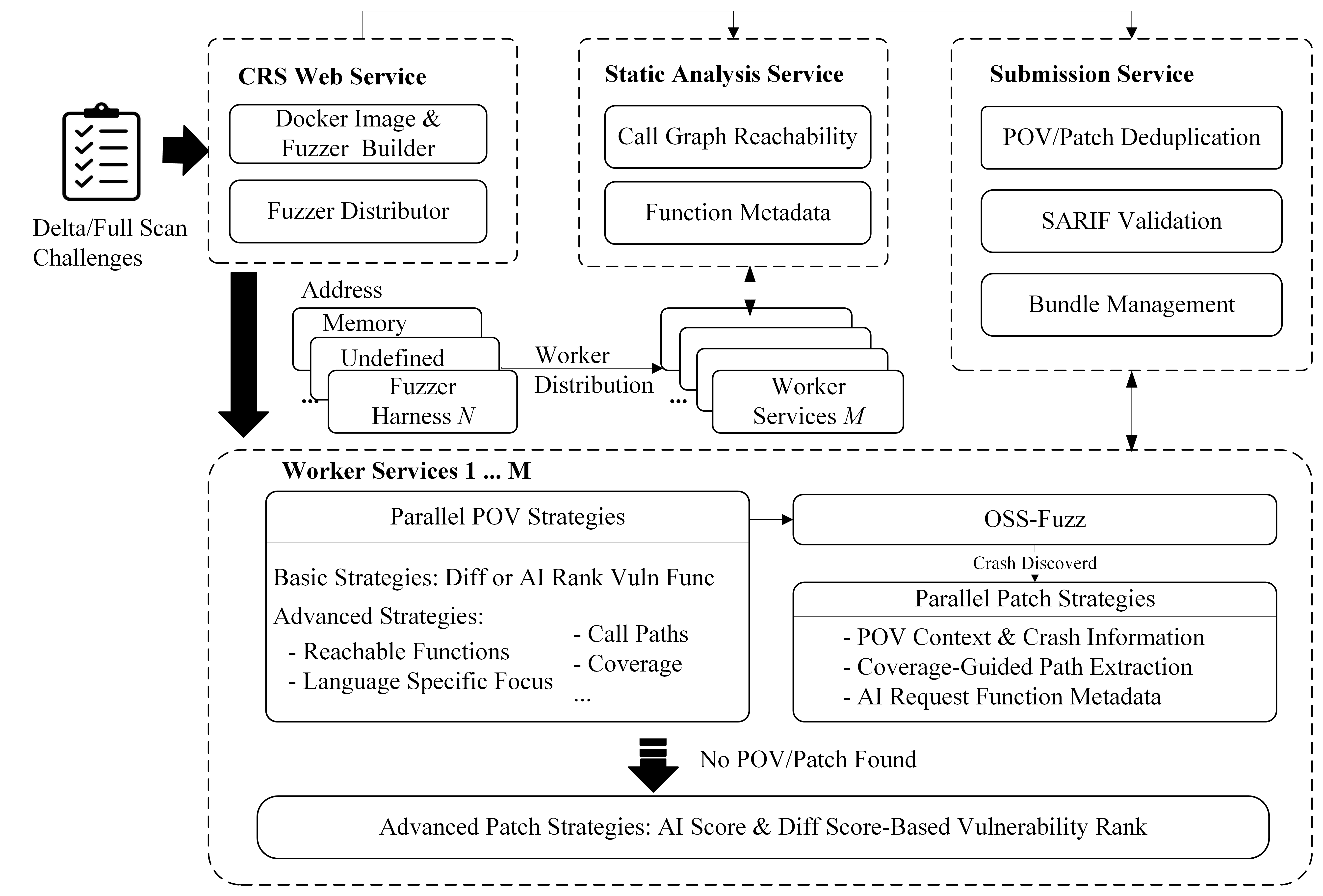} 
    \caption{Overview of  FuzzingBrain Architecture.}
    \label{fig:framework}
\end{figure*}

As depicted in Figure~\ref{fig:framework},
\textit{FuzzingBrain} consists of four core services: {\it CRS WebService}, {\it Static Analysis Service}, {\it Submission Service}, and {\it Worker Services}.  
All services execute in parallel on separate VM nodes. The first three run as single instances, while multiple instances of the {\it Worker Services} are deployed  (around 100 VMs in the final round) to support parallel task execution.

The \textbf{CRS Web Service} acts as the central coordinator. It decomposes tasks, builds fuzzers, and assigns them to worker services.  
The \textbf{Static Analysis Service} performs static code analyses to answer queries related to function metadata, reachability, and call paths.
The \textbf{Worker Services} generate POVs and patches by running fuzzing and LLM-based strategies. 
The \textbf{Submission Service} interacts with the competition API, handling submission deduplication, SARIF validation, and bundles. 

\subsection{Task Decomposition}
Figure~\ref{fig:crs} illustrates the workflow for decomposing and distributing tasks across FuzzingBrain’s services.  
Upon receiving a task (which contains metadata describing the challenge, e.g., a delta-scan or a full-scan of a target project), the {\em CRS Web Service} first builds the target project and its fuzzers using OSS-Fuzz utilities. Each fuzzer is an executable binary generated from a fuzzer harness instrumented with a sanitizer (AddressSanitizer, MemorySanitizer, or UndefinedBehaviorSanitizer for C/C++\footnote{FuzzingBrain also supports C++, besides C and Java projects.}, and
Jazzer for Java).  

The service then prepares isolated workspaces by cloning the target repository, constructing project-specific Docker containers, and generating fuzzer binaries for all supported sanitizer configurations. A single project may yield dozens of fuzzers: for example, {\tt dropbear} contains 17 fuzzer harnesses, each compiled with three sanitizers, producing more than 50 binaries.  

All fuzzers are distributed across the Worker Services to enable concurrent execution. To minimize communication overhead, only the fuzzer path (fuzzer name and sanitizer type) is sent to a worker. The corresponding binary and Docker image are reconstructed locally on the worker node.  

In parallel, the task is also dispatched to the Static Analysis Service and the Submission Service. The former conducts program analysis, while the latter manages bookkeeping for POV and patch submissions.  

\subsection{Libfuzzer and LLM-based Fuzzing}
Upon receiving a target fuzzer, a \textbf{Worker Service} first builds the corresponding fuzzer binary, and then proceeds to generate POV inputs and patches specific to that fuzzer. 
Both traditional fuzzing and LLM-powered fuzzing are performed on the same file system to discover inputs that can trigger sanitizer errors. Each fuzzer executes inside a Docker container with all required runtime dependencies.

The traditional fuzzing setup is intentionally minimal: we rely solely on \texttt{libFuzzer}, whose fuzzing corpus is configured to reside in a shared directory accessible by the LLM-based fuzzing strategies. All LLM-based strategies execute in parallel, and the inputs they generate that do not immediately trigger crashes are preserved in the shared corpus. These inputs are often close to valid crash-inducing cases and therefore serve as valuable seeds for \texttt{libFuzzer}. 
Since the parallel strategies can generate a very large number of test inputs per second, the shared corpus directory is periodically cleaned to prevent uncontrolled growth. 
Rather than performing libFuzzer's built-in corpus minimization, our approach removes files based on age: 
any input older than 10 minutes is deleted. 
This lightweight policy keeps the corpus size manageable while still allowing recently generated inputs (which are more likely to be relevant) to contribute to subsequent fuzzing iterations.  

For each LLM-based fuzzing strategy, the Worker Service spawns a dedicated Python subprocess. Once a POV is identified, the workflow transitions to the patching phase, where multiple patching processes are launched in parallel.  

\begin{table*}[htbp]
\centering
\caption{FuzzingBrain LLM-based Strategies}
\label{tab:strategies}
\begin{tabular}{|l|l|l|l|}
\hline
\textbf{Strategy Name} & \textbf{Mode} & \textbf{Language Focus} & \textbf{Stage} \\
\hline
\multicolumn{4}{|l|}{\textbf{Delta-Scan Strategies}} \\
\hline
xs0\_delta & delta-scan & C/C++, Java & POV Generation \\
as0\_delta & delta-scan & C/C++, Java & POV Generation \\
patch\_delta & delta-scan & C/C++, Java & Patch Generation \\
patch0\_delta & delta-scan & C/C++, Java & Patch Generation \\
patch1\_delta & delta-scan & C/C++, Java & Patch Generation \\
patch2\_delta & delta-scan & C/C++, Java & Patch Generation \\
patch3\_delta & delta-scan & C/C++, Java & Patch Generation \\
xpatch\_delta & delta-scan & C/C++, Java & Patch Generation \\
\hline
\multicolumn{4}{|l|}{\textbf{Full-Scan Strategies}} \\
\hline
xs0\_c\_full & full-scan & C/C++ & POV Generation \\
xs0\_java\_full & full-scan & Java & POV Generation \\
xs1\_c\_full & full-scan & C/C++ & POV Generation \\
xs1\_java\_full & full-scan & Java & POV Generation \\
xs2\_java\_full & full-scan & Java & POV Generation \\
as0\_full & full-scan & C/C++, Java & POV Generation \\
patch\_full & full-scan & C/C++, Java & Patch Generation \\
patch0\_full & full-scan & C/C++, Java & Patch Generation \\
patch1\_full & full-scan & C/C++, Java & Patch Generation \\
patch2\_full & full-scan & C/C++, Java & Patch Generation \\
patch3\_full & full-scan & C/C++, Java & Patch Generation \\
xpatch\_full & full-scan & C/C++, Java & Patch Generation \\
\hline
\multicolumn{4}{|l|}{\textbf{Report-Based Strategies}} \\
\hline
sarif\_POV0 & report-based & C/C++, Java & POV Generation \\
xpatch\_sarif & report-based & C/C++, Java & Patch Generation \\
\hline
\multicolumn{4}{|l|}{\textbf{Unharnessed Strategies}} \\
\hline
generate\_fuzzer & unharnessed & C/C++, Java & POV Generation \\
\hline
\end{tabular}
\end{table*}

FuzzingBrain currently incorporates \textbf{23 distinct LLM-based strategies} (10 for POV and 13 for patches), categorized by their operational mode and target language focus, as summarized in Table~\ref{tab:strategies}. Each strategy executes in an independent process but adheres to a unified interface for receiving task specifications and returning results in a consistent format. 
Further details of our LLM-based POV generation and patching strategies are presented in Sections~\ref{sec:pov} and~\ref{sec:patch}, respectively.

\subsection{XPatch without POV}
For certain complex challenges, generating POVs may be infeasible within the competition timeframe. 
To handle such cases, we developed {\bf XPatch}, a strategy that attempts to produce patches even when no POV has been found. 
XPatch is triggered only after half of the competition time has elapsed without a successful POV. 
According to the competition rules, such patches can still earn credit as long as they remediate the introduced vulnerability and do not regress against any known POVs. 
A detailed discussion of XPatch is provided in Section~\ref{ssec:xpatch}.  


\subsection{Static Analyses}
Upon receiving a task, the \textbf{Static Analysis Service} performs whole-program static analysis of the target project and supports three types of queries from Worker Services:  
(1) \emph{Function Metadata}---given a function name and an optional file name or path, return all matching functions along with their metadata, including parameters and source code;  
(2) \emph{Reachability}---given a fuzzer harness, identify all functions reachable from its entrypoint, returning each function’s name, file path, and start/end line numbers;  
(3) \emph{Call Paths}---given a fuzzing  harness and a target function, enumerate call paths from the fuzzer entrypoint to the target. Each call path comprises an ordered sequence of functions, including their file paths, names, and line ranges. To mitigate path explosion, we cap the number of call paths at 20, returning the first 20 if more exist. We also enforce a maximum call path depth (default: 50 for C/C++ and 10 for Java) to avoid excessively long paths.
These defaults were chosen heuristically based on empirical observations from the exhibition rounds: 
in C/C++ projects, vulnerabilities are often buried deep within complex call chains, making a higher threshold useful, 
whereas in Java projects, exploitable issues are typically exposed through shorter, higher-level entrypoints, so a smaller depth bound is sufficient in practice.  

Developing static analysis tools for real-world projects proved to be both challenging and time-consuming. We encountered numerous performance and soundness issues, ultimately leading us to build two customized static analysis frameworks: one for C/C++ and one for Java.
We present further details in Section~\ref{sec:staticanalysis}.

\subsection{Submission Deduplication}
The \textbf{Submission Service} receives all POV and patch submissions from Worker Services and applies several mechanisms to eliminate duplicates. Deduplication is essential because redundant submissions reduce the accuracy multiplier.

For POV submissions, each entry is accompanied by a \emph{signature}, defined as the crash location (source file and line number) extracted from the crash call stack. If the crash location is unavailable, heuristics are applied to construct a signature from the crash output and sanitizer. Two POVs are considered duplicates if they share the same signature. However, distinct signatures may still correspond to the same underlying vulnerability. To capture such cases, we employ LLM-based comparison: crash reports from two POVs are provided as input to three different LLMs, and the submissions are marked as duplicates if at least two of the models consider them redundant.

Patch submissions require a different strategy, since multiple distinct patches may be valid attempts for the same vulnerability, and some patches may fail during validation. Deduplication is therefore applied more conservatively, using three rules:  
(1) If two patch diffs are highly similar, we compute their Levenshtein distance and discard duplicates with a distance below 10. (2) If two patches are submitted within a short interval (3 seconds) and correspond to the same POV signature, the second of the two patches is discarded. (3) We cap the number of patch submissions per vulnerability at five.

For XPatch, where no POV is available, we assume that each task corresponds to a single introduced vulnerability. 
In this case, the canonical signature is derived from the task itself, and the submission cap is stricter: at most three XPatches are allowed per task.  



\subsection{SARIF Assessment}
For each SARIF broadcast, FuzzingBrain performs validation through LLM-based assessment and leverages the information for POV generation when appropriate. If a SARIF report is deemed valid and no POV has yet been discovered for the corresponding vulnerability, the CRS Web Service forwards the SARIF to Worker Services to guide POV generation. We present further details in Section~\ref{sec:sarif}.

\subsection{Bundle Creation}
Each bundle groups together two or more items (POV, patch, and/or SARIF broadcast) that correspond to the same underlying vulnerability. In our approach, to enable consistent grouping, we associate each vulnerability with a \emph{canonical signature}, defined as the signature of the first submitted POV. Patch submissions may also include an optional {\tt pov\_signature} when the patch is explicitly derived from a known POV.

Within the Submission Service, bundles are created and updated according to the following rules:  
\begin{itemize}
 \item POV submissions. Upon receiving a POV, if it is not marked as a duplicate and its status is \texttt{passed} (as confirmed by the competition API), then:  

   \begin{itemize}
   \item If the POV matches a SARIF that has been assessed as a true positive, a new bundle containing both the POV and SARIF is created.  
   \item  Otherwise, the POV initializes a bundle on its own. 
   \end{itemize}

 \item Patch submissions. Upon receiving a patch, if it is not marked as a duplicate and its status is \texttt{passed}, then:  
 
   \begin{itemize}
   \item If the patch shares a {\tt pov\_signature} with an existing POV, the patch is bundled with that POV.  
   \item If the POV is already in a bundle with a SARIF, the patch is added to the existing bundle, extending it to include all three components.  
   \end{itemize}

 \item SARIF broadcasts. Upon receiving a SARIF that has been validated as a true positive:  
   \begin{itemize}
   \item If the SARIF matches an existing POV, it is bundled together with that POV.  
   \item If the POV already belongs to a bundle (e.g., with a patch), the SARIF is added to that bundle.  
   \end{itemize}
   
\end{itemize}

\subsection{Technology Stack}
\begin{figure*}[htbp]
    \centering
    \includegraphics[width=1.05\textwidth]{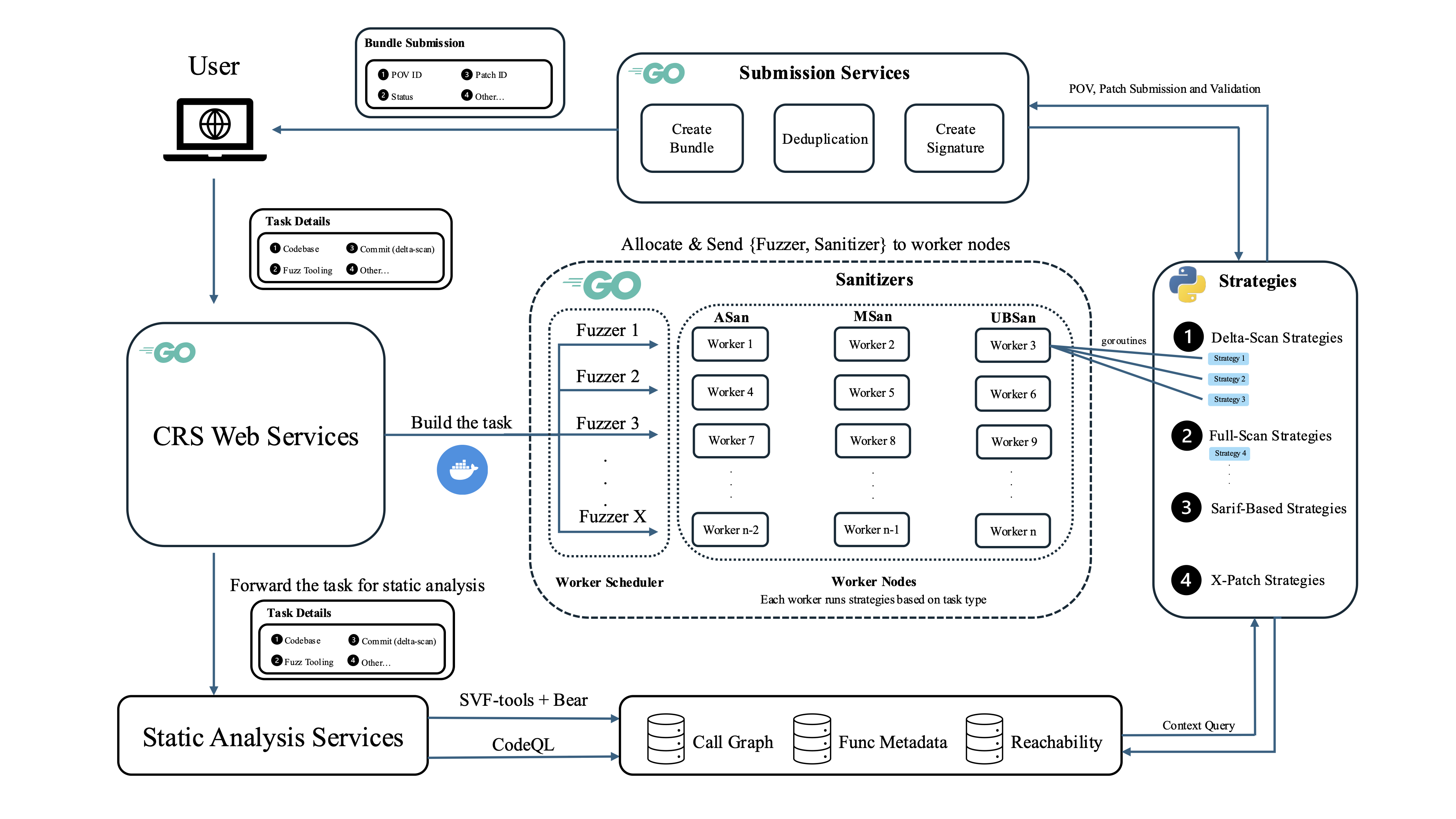} 
    \caption{Task Distribution \& Strategy Running}
    \label{fig:crs}
\end{figure*}

As illustrated in Figure~\ref{fig:crs}, FuzzingBrain is implemented primarily in two programming languages: Go and Python.  
For the CRS services, we selected Go with the Gin web framework. This choice reflects Go’s strengths in efficiently handling large numbers of concurrent operations and its mature ecosystem for building high-performance, production-grade web services.  

In contrast, all LLM-based POV and patching strategies are implemented as independent Python modules. Python was chosen due to its rich ecosystem for LLM development and its extensive set of third-party libraries, which allow rapid prototyping and flexible experimentation. Each strategy module can execute independently, enabling modularity and isolation.  

Finally, for model routing, we developed a custom framework for model selection, routing, and fallback. Existing off-the-shelf solutions were found to be unreliable and prone to errors under competition workloads (e.g., failing to handle high request volume, rate limits, server overloaded, etc). 
To enhance robustness against individual model failures or limitations, our framework employs a multi-model fallback mechanism. The system maintains a prioritized list of models, including those from Anthropic, Google, and OpenAI. When invoking an LLM, the framework attempts models in the predefined order; if one becomes unavailable or encounters an error, the request is automatically redirected to the next available model in the list. This design ensures continuity of service and minimizes disruptions during critical operations.

\subsection{Parallelization}
A central design principle of FuzzingBrain is \emph{parallelization}: every component that can be parallelized is parallelized, in order to maximize the speed of vulnerability discovery and patch generation.  

FuzzingBrain’s deployment infrastructure in the competition environment consists of:  
\begin{itemize}
  \item Approximately 100 virtual machines, each provisioned with 32–192 cores.  
  \item Each VM runs between 100 and 10,000 threads concurrently.  
\end{itemize}

This large-scale parallelization enables simultaneous processing of multiple challenges while maintaining high resource utilization and system throughput. The result is an architecture capable of scaling efficiently under heavy workloads, ensuring both rapid vulnerability detection and timely patch generation.

\section{LLM-based POV Strategies}
\label{sec:pov}

FuzzingBrain implements a total of 10 LLM-based POV generation strategies: two designed for delta-scans, six for full-scans, one for SARIF-based challenges, and one for unharnessed challenges (i.e., those without fuzzer harnesses and not scored).  

All strategies conform to a standardized framework built on iterative, dialogue-based interaction with LLMs. This feedback-driven refinement loop allows the system to incorporate execution results into successive iterations, enabling the LLM to learn from failed attempts and progressively improve its understanding of the target vulnerability.  

Each strategy executes as an independent process and adheres to a unified interface for task inputs and outputs. Strategies utilize five different frontier LLMs—\texttt{claude-3.7}, \texttt{chatgpt-latest} (gpt-4o at the time of the competition), \texttt{claude-opus-4}, \texttt{o3}, and \texttt{gemini-2.5-pro}. For each model, the framework performs multiple generation attempts, up to a maximum of five iterations by default. In addition, each strategy is subject to a configurable timeout (default: 30 minutes). 

\begin{figure*}[htbp]
    \centering
    \includegraphics[width=1\textwidth]{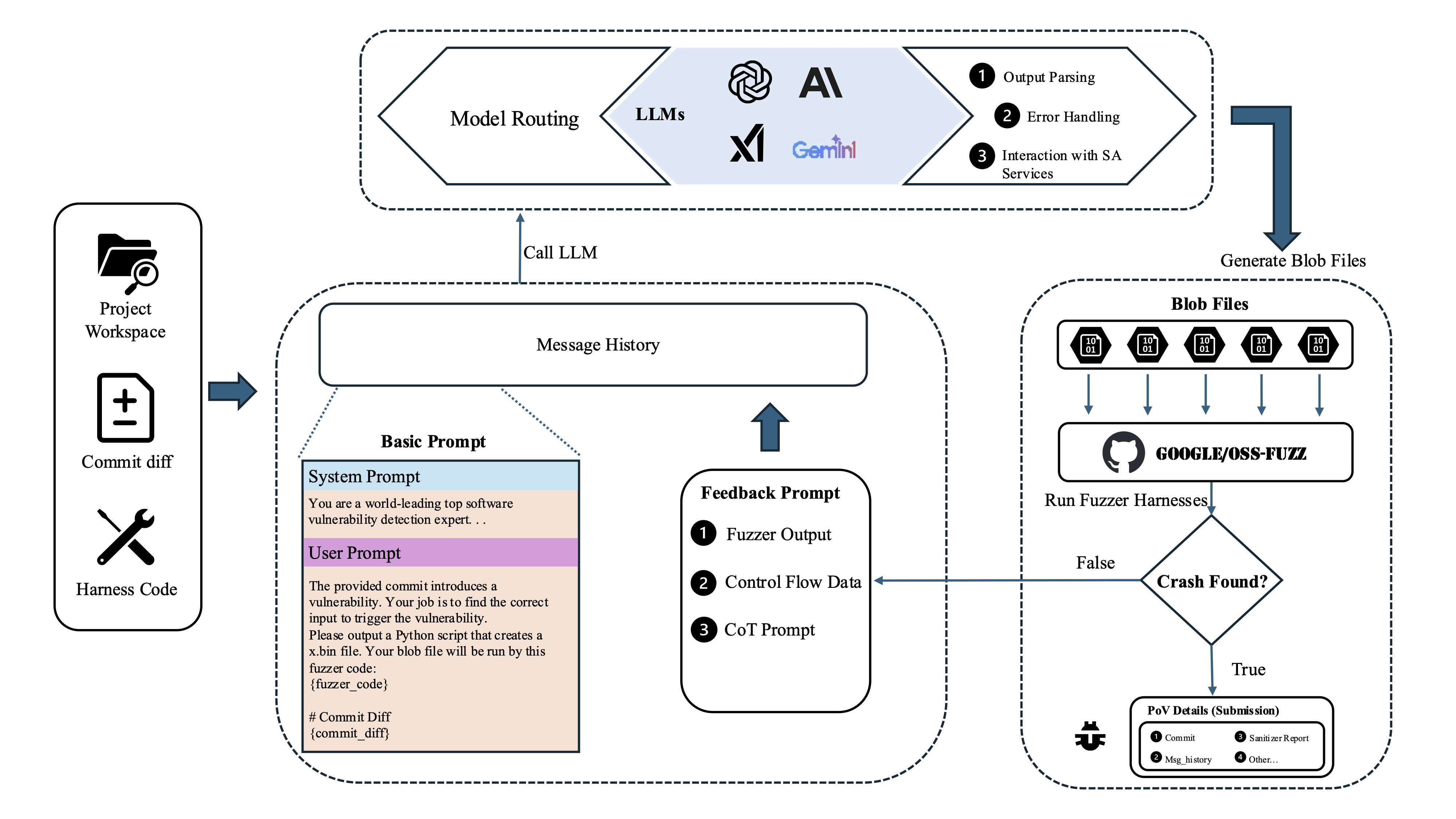} 
    \caption{Basic POV Generation Strategy}
    \label{fig:basic_strategy}
\end{figure*}

If a model fails to generate a valid POV within its iteration limit, the system automatically falls back to the next model in the priority list. This cascading mechanism increases the likelihood of success by leveraging the complementary strengths of different LLMs, as individual models often excel at distinct categories of vulnerability patterns.

\begin{table*}[htbp]
\centering
\caption{Delta-Scan POV Generation Strategies Comparison}
\label{tab:delta_POV_strategies}
\begin{tabular}{|l|l|l|l|}
\hline
\textbf{Strategy} & \textbf{Input Analysis} & \textbf{Generation Method} & \textbf{Core Characteristics} \\
\hline
\textbf{xs0\_delta} & Commit diff analysis & Single input per iteration & Basic strategy, iterative LLM refinement \\
\textbf{as0\_delta} & Commit diff analysis & \textbf{Multi-input generation} & Advanced strategy with 5 inputs per try \\
\hline
\end{tabular}
\end{table*}

\begin{table*}[htbp]
\centering
\caption{Full-Scan POV Generation Strategies Comparison}
\label{tab:full_POV_strategies}
\begin{tabular}{|l|l|l|l|}
\hline
\textbf{Strategy} & \textbf{Function Discovery} & \textbf{Ranking Method} & \textbf{Core Characteristics} \\
\hline
\textbf{xs0\_c\_full} & Call graph analysis & LLM ranking & Basic function filtering with simple LLM ranking \\
\textbf{xs0\_java\_full} & Call graph analysis & LLM ranking & Basic function filtering with simple LLM ranking \\
\textbf{xs1\_c\_full} & Dual reachability analysis & Multi-model LLM ranking & Parallel processing with improved function identification \\
\textbf{xs1\_java\_full} & Dual reachability analysis & Multi-model LLM ranking & Parallel processing with improved function identification \\
\textbf{xs2\_java\_full} & Dual reachability analysis & Multi-model LLM ranking & Advanced Java vulnerability detection with early termination \\
\textbf{as0\_full} & Call graph analysis & LLM ranking + fuzzing & Advanced generation with multi-phase strategies \\
\hline
\end{tabular}
\end{table*}

\subsection{Base Strategies}
FuzzingBrain implements a baseline strategy (\texttt{xs0\_delta}) for delta-scans, as illustrated in Figure~\ref{fig:basic_strategy}. This strategy establishes core prompting and feedback patterns that most other strategies inherit and extend.

The approach operates entirely via multi-turn, text-based dialogue with LLMs, cleanly decoupling analysis logic from the execution environment.

\paragraph{System Prompt.}
We initiate each dialogue with a high-level system instruction, for example:
\begin{quote}
\textit{``You are a world-class software vulnerability detection expert. Do not apologize when incorrect; instead, iteratively refine your analysis and proceed. When possible, identify any additional information that would improve your answer.''}
\end{quote}

\paragraph{User Message (Inputs).}
The user message combines the commit diff and the fuzzer harness source code, together with language- and sanitizer-specific guidance:
\begin{quote}
\textit{``The provided commit introduces a vulnerability. Your task is to craft an input for the fuzzer harness that triggers the vulnerability. Your input will be executed by the following harness: [\textbf{Fuzzer Source Code}] 
[\textbf{Sanitizer Guidance}] 
[\textbf{Language-Specific Guidance}] 
Your output must be a Python script that creates a file named exactly \texttt{x.bin}, and a short description of the vulnerability and the target function.''}
\end{quote}

\paragraph{Language- and Sanitizer-Specific Guidance.}
For C/C++ targets, the prompt includes sanitizer-focused instructions for AddressSanitizer (e.g., buffer overflows, use-after-free), MemorySanitizer (e.g., uninitialized reads), and Undefined-BehaviorSanitizer (e.g., integer overflows, pointer misalignment). For Java targets, we include Jazzer-oriented guidance (e.g., deserialization flaws, injection vectors, runtime exceptions).

\paragraph{LLM-Generated Python and Execution.}
We extract the Python code from the LLM response, execute it to generate \texttt{x.bin}, and then run the fuzzer harness on this input. 
At this stage, the harness is not performing fuzzing; it is simply used as an execution wrapper with sanitizers enabled to detect crashes (i.e., sanitizer errors). 
If a crash occurs, we record a successful POV and submit it to the Submission Service. 
If the attempt fails at any stage, the process continues with another LLM interaction, where the next user message provides structured feedback derived from the failure. For example:
\begin{quote}
\textit{%
``Fuzzer output: \{ \texttt{truncate\_output(fuzzer\_output, 200)} \} 
The test case did not trigger the vulnerability. Please analyze the output and try again. Consider: (1) alternative input formats/values; (2) edge cases; (3) focusing on functions modified in the commit; (4) careful attention to boundary conditions; (5) step-by-step reasoning.''}
\end{quote}

\paragraph{Coverage-Guided Feedback.}
If \texttt{x.bin} does not trigger a crash, we also supply coverage feedback in the next iteration. The feedback summarizes executed functions and branch decisions with $\pm$3 lines of surrounding source context.
For C/C++, we build with coverage instrumentation, execute the fuzzer on \texttt{x.bin} to produce \texttt{coverage.profdata}, then use \texttt{llvm-profdata} to derive an LCOV-style report from which we extract executed branches and nearby lines. For Java, we use JaCoCo to collect coverage data, generate a report, and post-process it to recover executed methods/branches with corresponding source excerpts.

\subsection{Advanced Strategies}

The \texttt{as0\_delta} strategy introduces several enhancements over the base approach to improve vulnerability discovery effectiveness. The primary differences among strategies are summarized in Table~\ref{tab:delta_POV_strategies} and Table~\ref{tab:full_POV_strategies}. In this section, we highlight advanced modules that extend the base strategy; different strategies compose these modules in different combinations.

\textbf{Multi-Input Generation.}
Unlike the base strategy, which produces a single test case per iteration, \texttt{as0\_delta} generates multiple candidates. Each LLM interaction emits a Python script that creates five binary inputs (\texttt{x1.bin}–\texttt{x5.bin}), yielding five exploitation opportunities per iteration instead of one.

\textbf{Vulnerability Category–Based Prompting.}
This module enumerates Common Weakness Enumeration (CWE) classes and applies category-specific prompts to guide input generation toward the intended weakness.  
\emph{C/C++ (10 categories):}
\begin{itemize}
\item CWE-119: Buffer Overflow
\item CWE-416: Use After Free  
\item CWE-476: NULL Pointer Dereference
\item CWE-190: Integer Overflow
\item CWE-122: Heap-based Buffer Overflow
\item CWE-787: Out-of-bounds Write
\item CWE-125: Out-of-bounds Read
\item CWE-134: Format String vulnerabilities
\item CWE-121: Stack-based Buffer Overflow
\item CWE-369: Divide by Zero
\end{itemize}
\emph{Java (representative examples; our implementation targets 15 categories):}
\begin{itemize}
\item CWE-22: Path Traversal
\item CWE-77/78: Command/OS Command Injection
\item CWE-79: Cross-Site Scripting
\item CWE-89: SQL Injection
\item CWE-502: Unsafe Deserialization
\item CWE-611: XML External Entity (XXE) Processing
\item CWE-918: Server-Side Request Forgery (SSRF)
\end{itemize}


\textbf{Modified-Function Context Injection.}
Beyond the commit diff, we identify all modified files and functions and append the full source of each modified function to the prompt to provide precise context. To stay within model context limits and emphasize salient code, we cap the injected source at 2{,}000 lines per function.

\textbf{Call-Path–Based Analysis.}
This module queries the Static Analysis Service for call paths from the fuzzer entrypoint to all modified (and thus potentially vulnerable) functions. For each call path, the system crafts a targeted prompt asking the LLM to generate an input that exercises that path, steering toward code regions likely to trigger the vulnerability. We limit the number of call paths to 20. If all per-path attempts fail, a final aggregated prompt combines all paths for one last POV-generation attempt.

\subsection{Full-Scan Strategies}
For full-scan scenarios, where no commit diff is available, FuzzingBrain employs a set of strategies that analyze the entire codebase to identify potentially vulnerable functions.  

\textbf{Call Graph–Based Analysis.}  
The \texttt{xs0\_c\_full} and \\ \texttt{xs0\_java\_full} strategies employ static analysis to narrow the search space prior to applying LLM-based vulnerability detection. Specifically, they query the Static Analysis Service to enumerate functions reachable from fuzzer entrypoints. This pruning step typically reduces the candidate set from thousands of functions to a more tractable subset of reachable targets.

\textbf{LLM-Based Vulnerable Function Ranking.}  
Once reachable functions are extracted, LLMs are used to score and rank them based on their likelihood of containing vulnerabilities. The ranking incorporates language-specific vulnerability patterns tailored to C/C++ and Java.  

\textbf{Enhanced Full-Scan Strategies.}  
The \texttt{xs1\_c\_full}, \\ \texttt{xs1\_java\_full}, and \texttt{xs2\_java\_full} strategies extend the baseline full-scan approach with more refined call graph construction, advanced ranking heuristics, and specialized vulnerability pattern recognition for their respective languages.  

\textbf{Advanced Full-Scan Integration.}  
The \texttt{as0\_full} strategy integrates all of the above modules (static call graph analysis, LLM-based ranking, and advanced vulnerability heuristics) into a unified workflow for large-scale vulnerability discovery across entire codebases.




\begin{figure*}[htbp]
    \centering
    \includegraphics[width=1\textwidth]{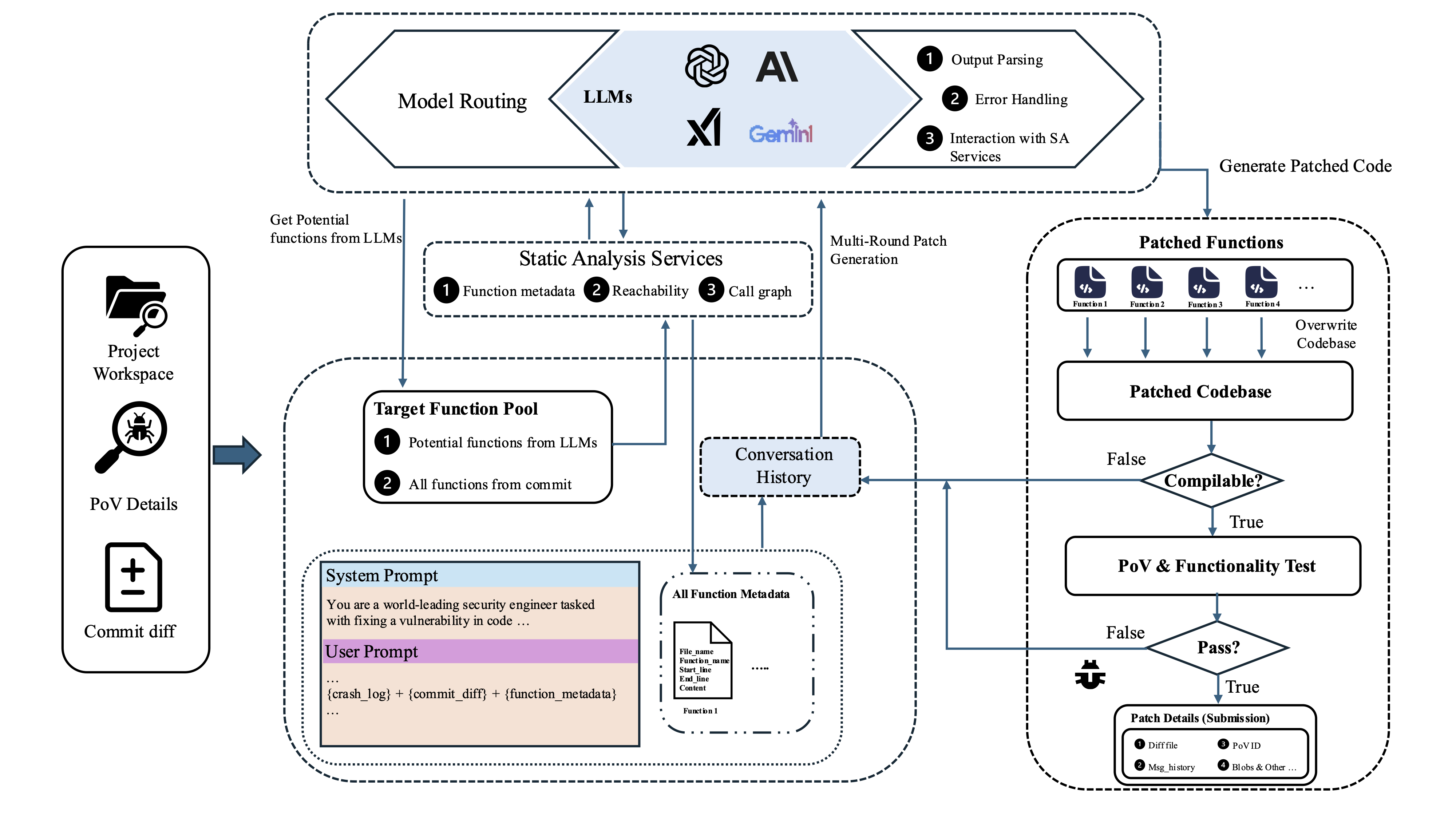} 
    \caption{Basic Patch Generation Strategy}
    \label{fig:patch_strategy}
\end{figure*}

\section{LLM-Based Patching Strategies}
\label{sec:patch}

FuzzingBrain implements 13 LLM-based patching strategies: six designed for delta-scans, six for full-scans, and one special XPatch strategy for generating patches without POVs. 

Except for XPatch, all patching strategies follow the same workflow, illustrated in Figure~\ref{fig:patch_strategy}:  
\begin{enumerate}
\item \textbf{Target Function Identification}: Identify vulnerable functions using strategy-specific heuristics.  
\item \textbf{Metadata Extraction}: Retrieve the complete function source code and surrounding context.  
\item \textbf{Patch Generation}: Use LLMs to produce a revised version of the function body.  
\item \textbf{Function Rewrite}: Replace the original function with the LLM-generated content.  
\item \textbf{Diff Creation}: Generate a \texttt{.diff} file using Git differential tools.  
\item \textbf{Validation}: Ensure compilation, execute POV tests, and run functionality tests.  
\item \textbf{Iterative Refinement}: Provide structured feedback for failed attempts, iterating until success or timeout.  
\end{enumerate}



\subsection{Patch Validation Criteria}
We define a patch as a code modification that mitigates a vulnerability without altering the program’s intended functionality. Patches are generated in standard \texttt{.diff} format and must satisfy four validation criteria:  
\begin{enumerate}
\item \textbf{Applicability}: The patch applies cleanly to the codebase.  
\item \textbf{Compilability}: The patched codebase compiles successfully.  
\item \textbf{Vulnerability Mitigation}: Known POVs (if available) no longer reproduce the vulnerability.  
\item \textbf{Functionality Preservation}: The patched codebase passes its functionality tests.  
\end{enumerate}
A patch is only considered valid if it passes all four criteria. 

\begin{table*}[htbp]
\centering
\caption{Delta-Scan Patch Generation Strategies Comparison}
\label{tab:patch_strategies}
\begin{tabular}{|l|l|l|l|}
\hline
\textbf{Strategy} & \textbf{Function Identification} & \textbf{Failure Feedback} & \textbf{Special Features} \\
\hline
\textbf{patch\_delta \& patch\_full} & LLM analysis only & Basic crash info & - \\
\textbf{patch0\_delta \& patch0\_full} & Diff extraction only & Basic crash info & - \\
\textbf{patch1\_delta \& patch1\_full} & LLM + Diff hybrid & Basic crash info & - \\
\textbf{patch2\_delta \& patch2\_full} & LLM + Diff & \textbf{+ Control flow paths} & Dynamic execution analysis \\
\textbf{patch3\_delta \& patch3\_full} & LLM + Diff & Basic crash info & Expert analysis + Sample patches \\
\hline
\end{tabular}
\end{table*}

\subsection{Basic Patch Strategy}
Table~\ref{tab:patch_strategies} shows a high-level comparison of the patching strategies. The baseline strategy, \texttt{patch\_delta}, operates through multi-turn LLM-driven conversations, similar to the POV generation process. In each iteration, the LLM proposes a candidate patch, which is validated against the criteria above. Successful patches are submitted; failures trigger detailed feedback and another iteration, up to the \texttt{MAX\_ITERATION} limit.

\textbf{Target Function Identification.}  
Target functions are identified as those suspected to be vulnerable (as determined by LLM analysis). This is the most critical step, as the quality of patching depends heavily on accurate function selection. Identification relies on structured prompts that combine commit diffs and crash logs, for example:  
\begin{quote}
\textit{"Your task is to identify all potentially vulnerable functions from a code commit and a crash log. The commit introduces a vulnerability. The vulnerability is found by an expert, with a crash log."}
\end{quote}

\textbf{Leveraging POV Generation Context.}  
This strategy reuses conversation history from the POV generation phase. The additional context, including prior failure cases and reasoning traces, improves the LLM’s ability to generate correct and targeted patches.  

\textbf{Function Metadata Extraction.}  
Once target functions are identified, the system queries the Static Analysis Service to extract detailed metadata, including function boundaries, complete source code, and precise file locations. This ensures the LLM has sufficient context to propose syntactically and semantically valid patches.  

\textbf{Multi-Model Resilience and Validation.}  
To improve reliability, the patching process employs multiple LLMs from different providers (e.g., Anthropic, Google, and OpenAI).
For each model, the system runs an iterative loop in which the LLM generates a candidate patch, the patch is applied to the codebase, and its validity is tested through compilation, execution of known POVs using the fuzzer harnesses (with sanitizers enabled), and, when available, functionality tests.  
If the patch fails at any stage, structured feedback—including compiler errors, failed diffs, or fuzzer output—is appended to the conversation, and the LLM attempts a revised patch in the next iteration. This process repeats until a valid patch is produced, the maximum iteration limit is reached, or the timeout expires. If one model fails to produce a valid patch, the system automatically falls back to the next model in the priority list.

\subsection{Greedy Strategy}

The \texttt{patch0\_delta} strategy implements a greedy approach that assumes vulnerable functions are guaranteed to appear within the commit diff.  

\textbf{Function Identification Optimization.}  
Rather than relying on LLM analysis, this strategy directly treats all functions present in the commit diff as vulnerable targets:  
\begin{quote}
\textit{Vulnerable Functions = all modified functions}
\end{quote}  

By extracting modified functions directly from the diff, this strategy reduces the computational overhead of LLM-driven identification in scenarios where vulnerabilities are clearly introduced through recent code changes.  

\textbf{Hybrid Enhancement.}  
The \texttt{patch1\_delta} strategy combines diff-based extraction with LLM analysis:  
\begin{quote}
\textit{Potential Vulnerable Functions = LLM-identified functions + all modified functions in the diff}
\end{quote}  
This hybrid design preserves the efficiency of direct diff extraction while broadening coverage with LLM-derived insights.  

\subsection{Path-Aware Strategy}

The \texttt{patch2\_delta} strategy enhances patch generation through dynamic execution analysis and enriched prompting.  

\textbf{Control-Flow Integration.}  
When patches fail validation, runtime control-flow data is incorporated into subsequent LLM prompts, similar to feedback mechanisms in POV generation. For C/C++ projects, coverage data is collected using LLVM profiling; for Java projects, JVM bytecode coverage analysis is employed. This information is embedded in prompts to enable path-aware patch generation.  

\textbf{Enhanced Function Discovery.}  
LLM prompting is extended to encourage the inclusion of functions beyond those explicitly mentioned in crash traces. For example:  
\begin{quote}
\textit{"You should include all functions that are potentially vulnerable, including not only those that appear in the crash call stack, but also those not directly mentioned in the crash log."}
\end{quote}  
This approach increases the likelihood of addressing indirect vulnerabilities and related security issues.  

\subsection{Knowledge-Enhanced Strategy}

The \texttt{patch3\_delta} strategy augments patch generation with expert vulnerability analysis and a curated patch catalog.  

\textbf{Expert Analysis Integration.}  
The strategy incorporates the initial analysis response from the POV generation phase as an expert assessment, providing contextual understanding of the vulnerability's characteristics and exploitation patterns.

\textbf{Sample Patch Catalog.}  
The strategy retrieves vulnerability-specific patch examples from a catalog indexed by sanitizer signatures and vulnerability categories. These examples serve as concrete guidance for remediation approaches.  

\textbf{Context Retrieval.}  
The strategy also supports dynamic context retrieval, enabling the LLM to request additional source code when needed:  
\begin{quote}
\textit{"If you need the source code of any other function, please return the file paths and function names in the following JSON format:"}
\end{quote}  
This feature supports comprehensive reasoning over complex vulnerabilities that span multiple functions or require broader program context.  

\subsection{Full-Scan Patch Strategy}
The main difference between full-scan patching strategies and delta-scan is the absence of a commit-based context, which forces function identification to rely solely on crash log analysis and LLM reasoning. To compensate, full-scan strategies employ enhanced prompting for more accurate target selection.  
Specifically, specialized prompts guide the LLM in analyzing beyond the immediate crash stack trace, encouraging exploration of indirectly related functions that may also contain vulnerabilities.   

\subsection{XPatch Strategy}
\label{ssec:xpatch}

The XPatch strategy addresses cases where no POVs or crash logs are available.  

For \textbf{delta-scans}, this process is relatively straightforward because the commit diff provides strong contextual clues about where the vulnerability was introduced. We extract all modified functions from the diff and supply their full source code as context to the LLM.  

For \textbf{full-scans}, where no commit information is available, XPatch relies on LLM-based scoring of all fuzzer-reachable functions. The LLM assigns likelihood scores to candidate functions based on predefined rubrics, and the top $k$ functions (default: 5) are saved. These functions are then used as the input context for patch generation.  

\textbf{Function Scoring Prompts.}  
The scoring prompt is tailored to both language and vulnerability class.  

  \begin{itemize}
\item For \emph{C/C++}, the rubric targets memory safety issues such as off-by-one errors, integer overflows, and buffer boundary violations.
Functions are scored from 1–10 to reflect the likelihood of a flaw: 
10 indicates a certain violation, 7–9 strong indicators, 2–6 weak or indirect hints, and 1 no evidence of problems.  

\item For \emph{Java}, multiple specialized rubrics are supported:  
  \begin{itemize}
    \item \emph{Malicious logic detection}: identifies intentionally harmful constructs such as backdoors, command injection, data exfiltration, privilege escalation, or kill-switches. Scores range from 1–10, where 10 indicates definite evidence of malicious intent, 7–9 strong indicators, 2–6 weak or indirect hints, and 1 no evidence of malicious behavior.  
    \item \emph{Unsafe deserialization}: flags dangerous uses of Java deserialization APIs without proper validation (e.g., unfiltered use of \texttt{ObjectInputStream}, XMLDecoder, or SnakeYAML). Scores range from 1-10 (similar to above).  
  \end{itemize}  
\end{itemize}  

In all cases, the LLM outputs a JSON array containing function names, assigned scores, and short justifications, sorted by descending score. Only functions with scores $\geq 7$ are retained.  

\textbf{Patch Generation.}  
Once candidate functions are identified, their metadata and source code are provided to the LLM, with explicit instructions that the vulnerability lies within one or more of these functions. The LLM is then tasked with generating candidate patches to mitigate the issue.  

\textbf{Patch Validation with LibFuzzer.}  
Since XPatch operates without known POVs, validation relies entirely on fuzzing. After applying a candidate patch, we execute LibFuzzer on the patched binary for 60 seconds. If the fuzzer produces a new crash during this run, the patch is deemed unsuccessful. Otherwise, the patch is considered valid under the available test conditions.  

\section{SARIF Analysis and Assessment}\label{sec:sarif}
\begin{figure*}[htbp]
    \centering
    \includegraphics[width=1\textwidth]{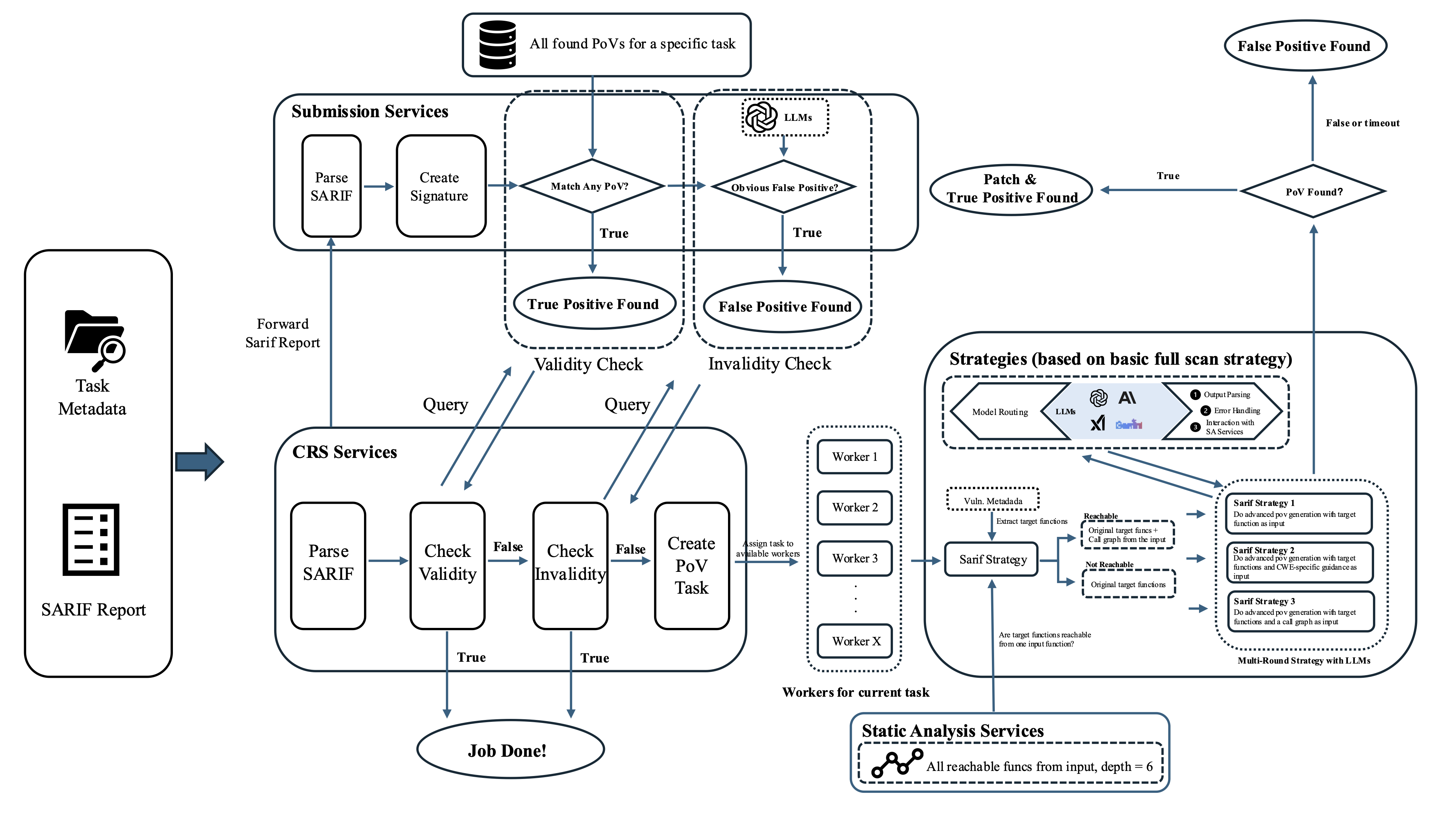} 
    \caption{SARIF Report-based Strategy}
    \label{fig:sarif_strategy}
\end{figure*}


Our SARIF (Static Analysis Results Interchange Format) report-based analysis implements a multi-stage processing pipeline that transforms external vulnerability reports into actionable security assessments. Figure \ref{fig:sarif_strategy} illustrates the workflow.

Upon receiving a SARIF broadcast, the CRS Web Service first parses the data to extract vulnerability metadata, including stack trace information, vulnerability classifications, precise code locations, affected functions, and contextual data. After processing, a typical SARIF report yields structured vulnerability information containing:
\begin{itemize}
\item \textbf{Affected Functions}: Target function names and their associated source file locations
\item \textbf{Vulnerability Classifications}: CWE identifiers and rule-based categorizations
\item \textbf{Location Information}: Precise line numbers, file paths, and code regions
\item \textbf{Contextual Metadata}: Severity levels, confidence scores, and analytical tool information
\item \textbf{Stack Trace Data}: Function call sequences and execution flow information
\end{itemize}

Based on this information, we identify potential vulnerable functions and use the Static Analysis Services to determine whether there is a fuzzer that can reach these vulnerable functions from its fuzzer input entry point. If not reachable, then we send a validation request to the Submission Service. This request includes relevant source code extracted from the target project, based on the vulnerable file, line numbers, and function name specified in the SARIF report (when available). At the Submission Service, three different LLMs are queried in sequence to perform two checks:  
1. Determine whether the SARIF is likely a false positive.  
2. Determine whether the SARIF is likely a true positive.  

If a majority consensus is reached, the resulting assessment is submitted to the competition API. If the outcome is inconclusive (e.g., conflicting results or LLM errors), the SARIF is deferred for later reassessment. Two additional mechanisms are then applied:  
1. When the Submission Service receives a POV, it checks whether the POV corresponds to any unprocessed SARIF reports. If a match is found, the SARIF is confirmed as valid.  
2. Independently, the SARIF is broadcast to Worker Services to drive POV generation. If a POV is successfully generated based on the SARIF, the SARIF is likewise confirmed as valid and submitted as such.  

Valid SARIF reports provide valuable contextual information, such as vulnerability description and precise source location, which can help improve the effectiveness of POV and patch generation. Accordingly, when a SARIF is classified as valid (but no POV has yet been identified) or remains undecided, it is forwarded to Worker Services to support further analysis and exploration.

The Submission Service handles SARIF validation because it tracks all POV submissions. This enables direct correlation between SARIF-reported vulnerabilities and actual POV-triggered crashes. The matching algorithm proceeds in two stages: (1) check whether the SARIF \texttt{artifactLocation} (e.g., vulnerable file and line numbers) appears in the POV crash trace, and (2) if necessary, use LLMs to compare the SARIF vulnerability description with the details of the POV submission.

\section{Static Analysis Implementation}\label{sec:staticanalysis}
\subsection{Static Analysis for C/C++}
Our C/C++ analysis pipeline integrates three external tools: LLVM~\cite{llvm}, SVF~\cite{svf}, and Bear~\cite{bear}.
The workflow is as follows: generate LLVM bitcode for each fuzzer binary, use SVF to construct a call graph, compute all functions reachable from each fuzzer, and then build call paths from the fuzzer entrypoint to each reachable function. To improve efficiency, all fuzzers are analyzed in parallel, and for each fuzzer–function pair, call paths are constructed concurrently using breadth-first search (BFS). The maximum call path depth is limited to 50.

Generating LLVM bitcode presented significant engineering difficulties. Simply appending {\tt -emit-llvm} to {\tt clang} often failed due to missing dependent headers. To address this, we first collect all compile commands of the target project into a compilation database by building it inside the OSS-Fuzz base Docker image via the command: {\tt bear -o /out/compile\_commands.json compile}.
This uses the Bear tool~\cite{bear}
%
%
to intercept and store the compilation command associated with each source file.
We then process each compile command entry to produce bitcode for each source file, maintaining separate bitcode sets for fuzzer harnesses and project source files. For each fuzzer, all bitcode files are subsequently linked into a single bitcode module. Despite this, errors persisted, requiring fallback heuristics such as hardcoding common header paths and compiler flags. Ultimately, our tool successfully generated bitcode for over 95\% of source files across tested projects, including \texttt{curl}, \texttt{dropbear}, and \texttt{sqlite3}.

Performance posed an additional challenge. Linked bitcode files could exceed 50~MB, and SVF frequently exceeded 1~hour or ran out of memory when constructing call graphs from such large bitcode files. To mitigate this, we trim oversized bitcode modules to a manageable size (e.g., 25~MB), apply lightweight type-based call graph analysis, and enforce a 10-minute timeout. If SVF fails to complete within this time budget, the Static Analysis Service will return empty results for that query.

\subsection{Static Analysis for Java}
For Java projects, we leverage CodeQL~\cite{codeql}, which provides a more mature and reliable analysis infrastructure than the toolchain used for C/C++. The workflow consists of three main steps: (1) building a CodeQL database from the project source code, (2) executing queries against the database, (3) decoding query results and computing per-fuzzer data structures (i.e., reachable functions and call paths). We implemented two custom queries: one for extracting reachable functions, and another for computing call paths. These queries are designed to handle challenges such as function overloading and dynamic loading, which complicate traditional static analysis approaches.

To maximize performance, multiple call path queries are batched to reduce database connection overhead (the primary latency bottleneck). The number of fuzzer–target pairs can be extremely large (up to 100K), so we employ a batch execution mode with a batch size of 1,000. To avoid race conditions, the database is cloned for each fuzzer path, ensuring that every fuzzer operates on an isolated copy. 

The call path analysis employs a balanced approach to cycle handling that prioritizes comprehensive path discovery over strict cycle prevention, that is, the code prevents direct cycles (intermediate methods cannot be the source or target method) but does not fully prevent indirect cycles. This design choice recognizes that indirect cycles often represent legitimate and valuable execution patterns in real world projects, such as recursive algorithms. The Analysis Service mitigates the theoretical risk of infinite recursion through a practical depth limit (10 calls), ensuring both computational efficiency and analytical completeness. 

In addition, we developed a lightweight baseline analysis that conservatively identifies function callees using only class types and function names. This approach sacrifices precision but serves as a fallback mechanism when CodeQL queries fail or become too costly.

Our Java static analysis pipeline completes within five minutes for representative OSS-Fuzz projects such as \texttt{Apache Zookeeper}, \texttt{Tika}, and \texttt{Commons-Compress}.

\section{Performance Optimizations}\label{sec:performance}
Across the three exhibition rounds, we observed that FuzzingBrain is effective and fast at producing POVs and patches: the vast majority were generated within the first 30 minutes. Most vulnerabilities were detected under AddressSanitizer, and our LLM-based strategies discovered nearly all POVs; traditional fuzzing (in our case, libFuzzer) contributed only one or two POVs. However, we also exhausted our allocated OpenAI API credits in one round. Postmortem analysis showed that a substantial portion of credits was spent on Worker Services assigned to fuzzers that could not reach the vulnerable code, making POV discovery impossible regardless of LLM effort.

Some target projects can yield more than 50 fuzzers (fuzzer harnesses $\times$ \{address, memory, undefined\} sanitizers). To curb waste and improve throughput for the final round, we applied the following policies:

\paragraph{Sanitizer Selection.}
We disabled UndefinedBehaviorSanitizer for all projects and disabled MemorySanitizer for projects with more than ten fuzzer harnesses, prioritizing AddressSanitizer where we observed the highest yield.

\paragraph{Time Budgeting for LLM-Based Fuzzing.}
To control API spend, each worker's LLM-based fuzzing is capped at 60 minutes, and reduced to 45 minutes if other fuzzers have already produced POVs for the same target.

\paragraph{libFuzzer Time Management.}
To preserve CPU for concurrent fuzzers on the same VM, we limit libFuzzer to at most half of the competition time. Concretely (as implemented in our controller):
\begin{itemize}
  \item If one or more POVs have been found (either by LLM-based strategies or libFuzzer) and libFuzzer has already run longer than the half-time budget (or longer than half that budget while multiple fuzzers are active on the same VM), we stop libFuzzer to save resources.
  \item If no POVs found, we continue until the half-time budget is reached, then stop.
\end{itemize}

\paragraph{Parallelism and Isolation.}
Because patching requires full rebuilds (often minutes per attempt), we run 3--5 parallel processes for each patching strategy. A single VM may execute 20--30 patching processes concurrently for one vulnerability. Each process operates in its own isolated workspace to prevent cross-contamination between attempts.

\paragraph{Patch Submission Caps per Vulnerability.}  
Submitting too many patches, even if valid, reduces the accuracy multiplier and lowers the overall score. Because patches contribute the largest share of points, we balance patching success probability against scoring penalties by capping the number of submissions per vulnerability (keyed by canonical signature): at most five POV-based patches and at most three XPatches.  

These treatments reflect tradeoffs between efficiency and coverage, aiming to optimize performance under resource constraints. However, they may not guarantee the best possible score in every setting. For example, if many vulnerabilities can only be triggered under MemorySanitizer or UndefinedBehaviorSanitizer, skipping these sanitizers would cause FuzzingBrain to miss them, leading to lower overall performance in the final round.  

\section{Additional Lessons Learned}\label{sec:lessons}


Engineering a reliable LLM-based system of this scale proved extremely challenging. We encountered countless bugs, spent many days debugging, and learned several important lessons across software engineering, infrastructure management, and system design.  

\subsection{Concurrency and Parallelization}
FuzzingBrain relies heavily on parallelism, which introduced subtle race conditions and deadlocks:
\begin{itemize}
  \item In one exhibition round, the Submission Service froze after a few hours. We later traced the root cause to a classical deadlock in Go mutex usage across multiple threads. The issue was resolved by removing mutexes and replacing shared maps with \texttt{sync.Map}.  
  \item Worker services occasionally submitted false-positive POVs due to file-level race conditions. Multiple POV strategies wrote to the same output path (\texttt{x.bin}), leading to mismatched files being submitted. The fix was to isolate file paths for each subprocess.  
\end{itemize}

\subsection{LLM-Generated Code and Debugging}
More than 90\% of our system code was generated with LLM assistance. While this accelerated development, it also made debugging significantly harder, as we were not as familiar with code we did not write ourselves. This experience highlighted the tradeoff between rapid prototyping and long-term maintainability when relying heavily on LLMs for code generation.  

\subsection{Configuration Management}
With multiple services (each requiring its own API keys), we maintained four separate \texttt{.env} files across subdirectories. In one exhibition round, a critical service failed because its file was not updated with the correct keys. This underscored the need for centralized and automated secret management, rather than relying on manual updates.  

\subsection{Logging and Observability}
According to AIxCC competition logs, our system performed extremely well for the first three days of the final round, but submitted nothing after Day 4. We suspect a critical crash or bug, but the root cause remains unknown due to loss of logs:
\begin{itemize}
   \item Our system generated gigabytes of logs across services within hours, but most logs were stored only on ephemeral VM nodes in the Azure VMSS cluster.  
  \item After the competition, all VM nodes were recycled, and the logs disappeared permanently, preventing postmortem analysis.  
\end{itemize}
This revealed the importance of persistent, centralized logging and monitoring in large-scale distributed systems.  

\subsection{Validation and Silent Failures}
Post-final analysis revealed a critical flaw in our patch validation pipeline: a missing parameter in a Python function call caused all Python subprocesses (invoked from Go to handle LLM-based patching) to crash silently. Consequently, many patches were submitted without verification against known POVs, leading to a significant drop in accuracy. This highlights the need for robust error handling, explicit status reporting, and fail-safe validation mechanisms when orchestrating heterogeneous components.

\subsection{Diverse Fuzzing Strategies}
Our system made only minimal use of traditional fuzzing (only libFuzzer). While LLM-based fuzzing was highly effective, we likely missed opportunities to discover additional POVs by not integrating alternative fuzzers such as AFL++~\cite{afl_plus_plus} or Honggfuzz~\cite{honggfuzz}. Incorporating diverse fuzzing engines could have improved overall coverage and robustness.  



\section{FuzzingBrain LeaderBoard}\label{sec:leaderboard}

To systematically evaluate state-of-the-art LLMs on vulnerability detection and patching, we developed the \emph{FuzzingBrain Leaderboard} based on the AIxCC benchmarks (36 challenges drawn from the three exhibition rounds, 16 C challenges and 20 Java challenges). In each run, FuzzingBrain is restricted to using a single LLM for both POV generation and patching, allowing us to directly measure the performance of that model. Scoring follows the AIxCC rubric: each POV is worth 2 points and each patch is worth 6 points. Models are then ranked according to their total score across all benchmarks, providing a standardized comparison of capability.  

We introduce several modifications to make leaderboard evaluation practical and reproducible:  
\begin{itemize}
  \item \textbf{Single-VM Execution:} FuzzingBrain is executed on a single VM, and the vulnerability-triggering fuzzer is provided as input.  
  \item \textbf{Precomputed Static Analysis:} Static analysis results for each target project are precomputed and stored in JSON format. The Static Analysis Service, therefore, only needs to answer queries and return results, minimizing runtime overhead.  
  \item \textbf{Time Limit:} Each run is capped at one hour in total, covering both POV generation and patching.  
\end{itemize}

The current leaderboard is available at \href{https://o2lab.github.io/FuzzingBrain-Leaderboard/}{\textcolor{blue!70!black}{o2lab.github.io/FuzzingBrain-Leaderboard}}. We plan to regularly maintain this evaluation framework to enable transparent, standardized, and reproducible comparisons of different LLMs in real-world vulnerability discovery and remediation tasks.

\bibliographystyle{ACM-Reference-Format}
\bibliography{refs}

\end{document}